\documentclass[twocolumn,showpacs,preprintnumbers,amsmath,amssymb]{revtex4}

\usepackage{graphicx,psfrag}
\usepackage{dcolumn}
\usepackage{bm}

\begin{document}

\preprint{APS/000}

\title{Two-time synchronism and induced synchronization }

\author{Przemys{\l}aw Szlachetka}
 \email{przems@main.amu.edu.pl}
\author{Krzysztof Grygiel}
 \email{grygielk@main.amu.edu.pl}
\author{Marcin Misiak}
 \email{misiak@zon10.physd.amu.edu.pl}
\affiliation{Nonlinear Optics Division, Institute of Physics,
A. Mickiewicz University,
 ul. Umultowska 85, PL 61-614 Pozna\'{n}, Poland }

\date{\today}

\begin{abstract}
A pair $({\bf A},{\bf B})$  of interacting oscillators treated as a master system sending
  signals to its slave copy $({\bf a},{\bf b})$ through  two
  communication channels
  ${\bf A} \Rightarrow {\bf a}$ and ${\bf B}\Rightarrow {\bf b}$ is considered. The effect of 
  non-simultaneous (two-time) synchronization of the pairs
  $({\bf a}{\bf ,A})$  and $({\bf b},{\bf B})$  is demonstrated with the help of coupled Kerr
  oscillators  producing hyperchaos. An individual  pair, for example, $({\bf
  b},{\bf B})$  can
 also be synchronized  when its  communication channel
  ${\bf B} \Rightarrow {\bf b}$ is turned off, provided that the second
  channel for the pair$({\bf
  a},{\bf A})$ is turned on.  The resulted 
  synchronization is termed  induced.  The efficiencies of the
  presented synchronization precesses are studied..
\end{abstract}

\pacs{ 05.45.Xt, 42.65.Sf}
                             
\maketitle

\begin{figure}
%\begin{center}
\includegraphics[width=8cm,height=5cm,angle=0]{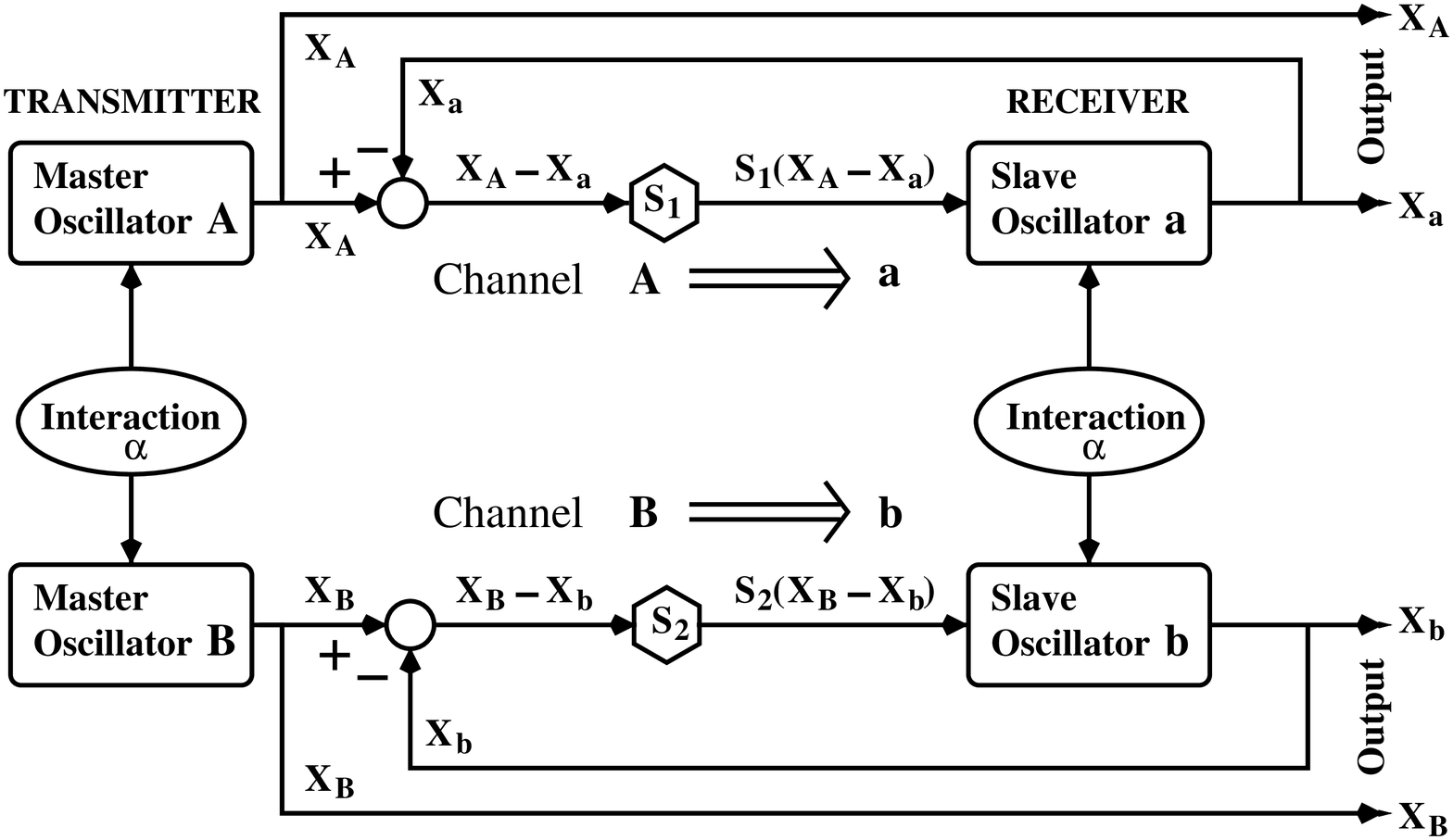} 
%\end{center}
\caption{Schematic diagram of synchronization. Signals from two
  interacting master oscillators ${\bf A}$ and ${\bf B}$ are sent to
  a system containing two corresponding slave oscillators ${\bf
  a}$ and ${\bf b}$. The signals are
  controlled by the parameters $S_{1}$ and $S_{2}$. 
  The synchronization state is achieved if $ X_{a}(t)=X_{A}(t)$ and 
$X_{b}(t^{'})=X_{B}(t^{'})$. 
The  question is whether and when $t=t^{'}$. } 
\label{Fig.1}
\end{figure}

\begin{figure}
\psfrag{q1}[b][]{\large $ q_{a}$}
\psfrag{q5}[b][]{\large $q_{A}$}
\psfrag{t}[t][]{\large $t$}
\begin{center}
\includegraphics[width=3cm,height=8cm,angle=-90]{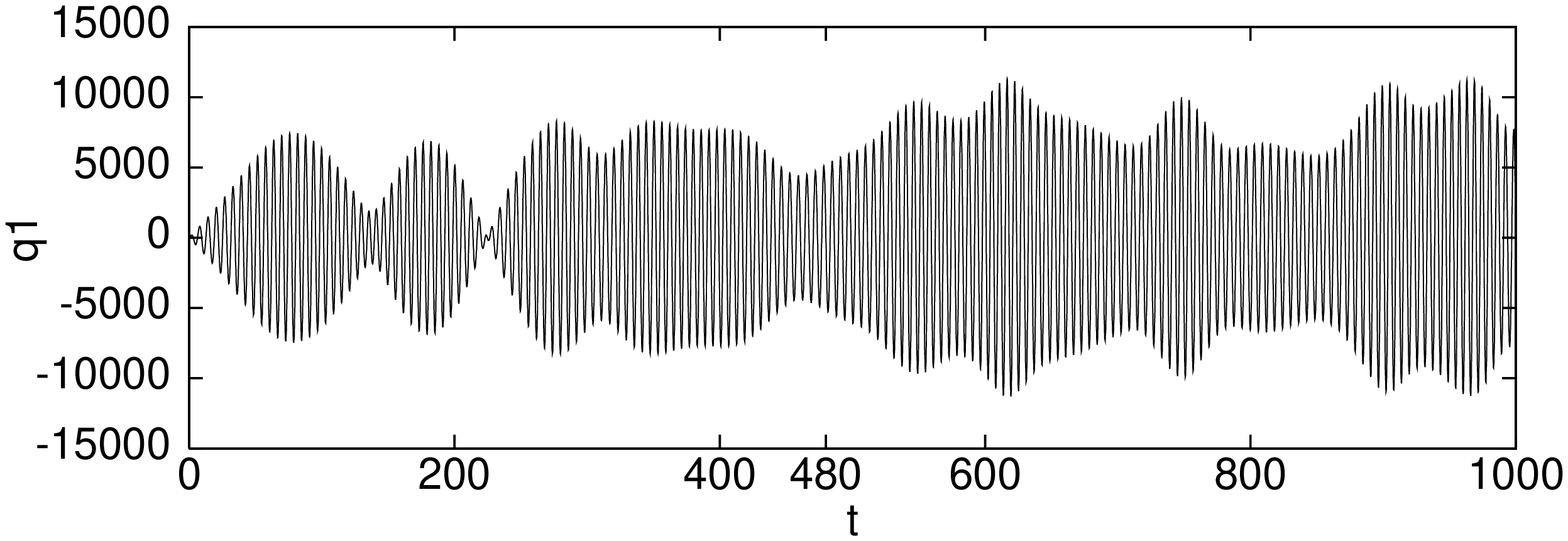} 
\includegraphics[width=3cm,height=8cm,angle=-90]{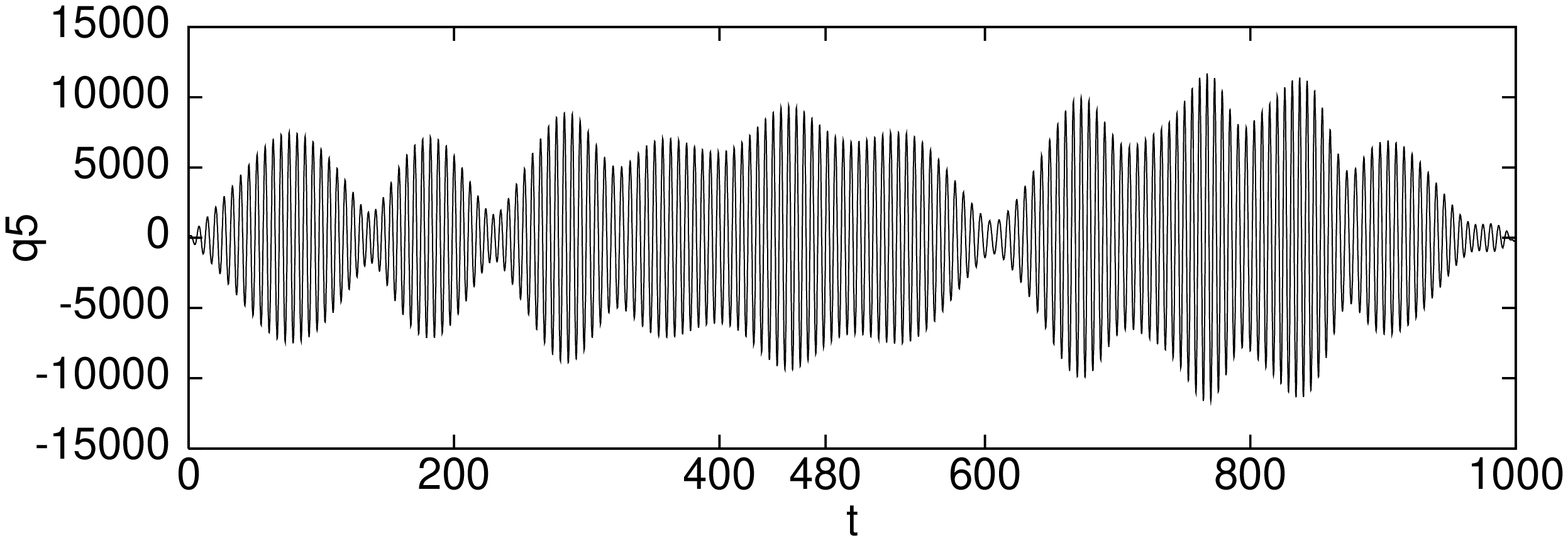} 
\end{center}
\caption{ Evolution of $q_{a}$  and $q_{A}$ vs $t$ 
 for Eqs. (\ref{2aa})--(\ref{2ds}) if $S_{1}=S_{2}=0$.} 
\label{Fig.2}
\end{figure}

 Recently, there has been a great deal of interest in the study of
 coupled  oscillators 
 and their role in explaining the
basic features of man-made and natural systems. Such systems can exhibit 
behaviors such as on-off
intermittency \cite{platt}, two-state on-off intermittency \cite{ying} or beats with chaotic
envelopes \cite{wiley}. In particular, much attention has been paid to
 synchronization of chaotic systems. Different types of
synchronization have been considered, for example, complete
synchronization \cite{pecora}--\cite{pecora2}, 
 partial synchronization \cite{Pyrygas,Ying},  generalized
 synchronization \cite{Rulkov}--\cite{Rulkov1} or phase
synchronization \cite{Pikovsky,Rosenblum}.
   Especially, 
  the problem of synchronization of coupled chaotic
oscillators has been intensively studied  mainly in view of
potential application to secure communication
 \cite{Cuomo}, \cite{Kocarev1}--\cite{Liao}. 
 The idea of synchronization has also been  implemented to higher
  dimensional systems  exhibiting hyperchaotic behavior \cite{Peng}-\cite{Ali}. 
This seems to be very impressive  as  multidimensional systems
 improve the degree of security in  communication. Quite recently, a
 scheme of  dual chaos synchronization has
been proposed \cite{Liu}. In the dual synchronization,  
signals from two noninteracting master
oscillators  through a single communication channel are sent to a system
containing two corresponding slave oscillators.\\
  In this short paper we  discuss the possibility of applying a chaos
 controlling method to achieve  synchronization of two different
 pairs of  oscillators. The general set up to be considered is
 presented in Fig. \ref{Fig.1} 
%%%%%%%%%%%%%%%%%%%%%%%%%%%%%%%%%%%%%% Figure 1 %%%%%%%%%%%%%%%%%%%%%%%%%%%%%%%%%%%%
%%%%%%%%%%%%%%%%%%%%%%%%%%%%%%%%%%%%%%%%%%%%%%%%%%%%%%%%%%%%%%%%%%%%%%%%%%%
The  master  system consists of two coupled oscillators $({\bf A},{\bf B})$ which
interact with each other ( the symbol
$\alpha$ denotes a parameter of interaction  between ${\bf A}$ and ${\bf B}$). 
If $\alpha=0$ the  master system consists of two independent oscillators. 
The slave  system $({\bf a},{\bf b})$ is a copy of the master system. The signals from
the two master subsystems $({\bf A},{\bf B})$ are  transmitted to the their
counterparts $({\bf a},{\bf b})$ in the slave
system by linear feedback functions.  The control  parameters are
 denoted by $S_{1} $ and $S_{2}$, respectively. 
The slave and masters systems are assumed to start from  different initial conditions.
  As a master system $({\bf A},{\bf B})$ let us consider two coupled Kerr  oscillators 
  governed by the following equations \cite{grygiel}:
\begin{eqnarray}
 \label{2aa}
 \frac{dq_{A}}{dt}&=&p_{A}[1+\epsilon(p_{A}^{2}+\omega^{2}q_{A}^{2})] -\gamma q_{A}\,,\\
 \label{2bb}
\frac{dp_{A}}{dt}&=&-\omega^{2}q_{A}[1+\epsilon
(p_{A}^{2}+\omega^{2}q_{A}^{2})]-\gamma p_{A}
\nonumber\\
&+&\alpha q_{B}+A\cos \Omega_{1}t \,,\\
 \label{2cc}
\frac{dq_{B}}{dt}&=&p_{B}[1+
\epsilon(p_{B}^{2}+\omega^{2}q_{B}^{2})]-\gamma q_{B}\,,\\
 \label{2dd}
\frac{dp_{B}}{dt}&=&-\omega^{2}q_{B}[1+\epsilon
(p_{B}^{2}+\omega^{2}q_{B}^{2})]-\gamma p_{B}
\nonumber\\ 
&+&\alpha q_{A}
+A\cos \Omega_{2}t\,, 
\end{eqnarray}
 where $p_{A,B}$ and $q_{A,B}$ are the momentum and position, respectively.
The anharmonic Kerr terms are  denoted by
$\epsilon(p_{A,B}^{2}+\omega^{2}q_{A,B}^{2})$, 
where  $\epsilon$ is the parameter of Kerr nonlinearity, and $\omega$
is the natural frequency of a simple harmonic oscillator.
 The individual Kerr
oscillators are pumped by  the external time-dependent forces 
with the amplitude $A$ and frequencies  $\Omega_{1}$
and $\Omega_{2}$, respectively. A loss mechanism is governed by the
terms $\gamma p_{A,B}$ and $\gamma q_{A,B}$.
 The $\alpha$-terms  are  responsible for the linear
interaction between the individual Kerr subsystems.  
Due to different frequencies  $\Omega_{1}$
and $\Omega_{2}$ the states of the systems ${\bf A}$ and ${\bf
  B}$ are not identical at any time i.e. $q_{A}(t)\neq q_{B}(t)$ and
$p_{A}(t) \neq p_{B}(t)$. If $\epsilon=0$, Eqs. (\ref{2aa})--(\ref{2dd})
describe a standard text-book model of two coupled linear subsystems.
The Kerr systems  occupy an important
position in optical  devices  and have been extensively
investigated  in the classical as well as  quantum approach (for a review, see
Ref.~\cite{gerls,Grygiel1}). 
 We study  Eqs.  
(\ref{2aa})-(\ref{2dd})  numerically  with the help of the
fourth-order Runge-Kutta method with the integration step $\Delta t=0.01$.
We fix the parameters at $\omega=1$ , $A=200$,
 $\epsilon=10^{-9}$, $\alpha=0.04$, $\gamma=0.001$, $\Omega_{1}=1$
 and $\Omega_{2}=1.05$. The system  starts from the initial conditions
$q_{A}(0)=100$, $p_{A}(0)=0$, $q_{B}(0)=100$ and $ p_{B}(0)=0$. 
 The spectrum of Lyapunov exponents for the system 
(\ref{2aa})--(\ref{2dd}) computed by the method of Wolf {\em et
  al.}\cite {runge} 
  is given by $\{0.008,0.004,-0.007,-0.001\}$, which means that the
  system is hyperchaotic.
 According to the continuous feedback method
~\cite{pecora,pyragas},
our master system $({\bf A},{\bf B})$ is coupled to the slave system
$({\bf a},{\bf b})$ in
the following way
%%%%%%%%%%%%%%%%%%%%%%% 
 \begin{eqnarray}
 \label{2as}
 \frac{dq_{a}}{dt}&=&p_{
a}[1+
 \epsilon (p_{a}^{2}+\omega^
{2}q_{a}^{2})]-\gamma q_{a}\nonumber\\
  &+&S_{1}(q_{A}-q_{a})\,,\\
 \label{2bs}
\frac{dp_{a}}{dt}&=&-\omega^{2}q_{a}[1+\epsilon
(p_{a}^{2}+\omega^{2}q_{a}^{2})]
 -\gamma p_{a}
\nonumber\\
&+&\alpha q_{b}+A\cos \Omega_{1}t  +S_{1}(p_{A}-p_{a})\,,\\
 \label{2cs}
\frac{dq_{b}}{dt}&=&p_{b}[1+
\epsilon (p_{b}^{2}+\omega^{2}q_{b}^{2})]-\gamma q_{b}\nonumber\\
&+&S_{2}(q_{B}-q_{b})\,,\\
 \label{2ds}
\frac{dp_{b}}{dt}&=&-\omega^{2}q_{b}[1+\epsilon(p_{b}^{2}+\omega^{2}q_{b}^{2})]-\gamma
p_{b}
\nonumber\\ 
&+&\alpha q_{a}+A\cos \Omega_{2}t+S_{2}(p_{B}-p_{b}) \,.
\end{eqnarray}
 The system (\ref{2as})-(\ref{2ds}) is examined  for  the same values of
the parameters as the master system but for  different initial conditions
which are fixed to be $ q_{a}(0)=1$, $p_{a}(0)=0$, $q_{b}(0)=1$ and 
$ p_{b}(0)=0$. 
If the feedback terms $(q_{A,B}-q_{a,b})$ and $(p_{A,B}-p_{a,b})$ in Eqs. (\ref{2as})--(\ref{2ds})
are switched off the master and slave  systems generate hyperchaotic
  beats  shown in Fig.\ref{Fig.2}.
%%%%%%%%%%%%%%%%%%%%%%% Fig 2 %%%%%%%%%%%%%%%%%%%%%%%%%%%%%%%%%%%%%%%%%%%%%%%%%%%
%%%%%%%%%%%%%%%%%%%%%%
  As seen, the beats whose  structure
  resembles the original trace voice {\cite{Short} have chaotic
  envelopes and stable fundamental frequencies
  \cite{wiley,Grygiel1}.
Let us now consider  the synchronization of beats presented in Fig.\ref{Fig.2}, 
when  the feedback terms are switched on at the time $ t_{0}=480.$ The
choice of the initial time $t_{0}=480$ is motivated by the fact that
for this time the states of the master and drive systems are
distinctly different as is clearly seen in Fig.2. 
The synchronization time $T_{s}$ is defined as $T_{s}=t_{s}-t_{0}$,
  where $t_{s}$ is defined as the time after which
the quantity  $q_{A,B}-q_{a,b}$ is less than $10^{-3}$.
 The dynamics of synchronization is strongly different  for the cases $S_{1}\neq S_{2}$  and
  $S_{1}= S_{2}$. For  $S_{1}= S_{2}$ the time of
  synchronization for the pairs of oscillators
  $({\bf a},{\bf A})$ and $({\bf b},{\bf B})$ remains  the same.
 However, for the control parameter $S_{1}\neq S_{2}$  
 we can observe two  different times of synchronization $T_{s}^{(a,A)}$ and $T_{s}^{(b,B)}$  for the
  master-drive subsystems $({\bf a},{\bf A})$ and $({\bf b},{\bf
  B})$, respectively. This two-time synchronism is illustrated in Fig.\ref{Fig.3} for
 the values  $S_{1}=5$ and $S_{2}=0.025$. 
%%%%%%%%%%%%%%%%%%%%%%%%%%%%%%%%%%%%%% Figure 3 %%%%%%%%%%%%%%%%%%%
\begin{figure}
\psfrag{q1q5}[t][]{\large $\Delta_{(a,A)}$}
\psfrag{q3q7}[t][]{\large $\Delta_{(b,B)}$}
\psfrag{t}[l][]{\large $t$}
\psfrag{ts=409}[t]{$T_{s}^{(a,A)}=409$ }
\psfrag{ts=790}[t] {$T_{s}^{(b,B)}=790$}
\begin{center}
\includegraphics[width=5cm,height=8cm,angle=-90]{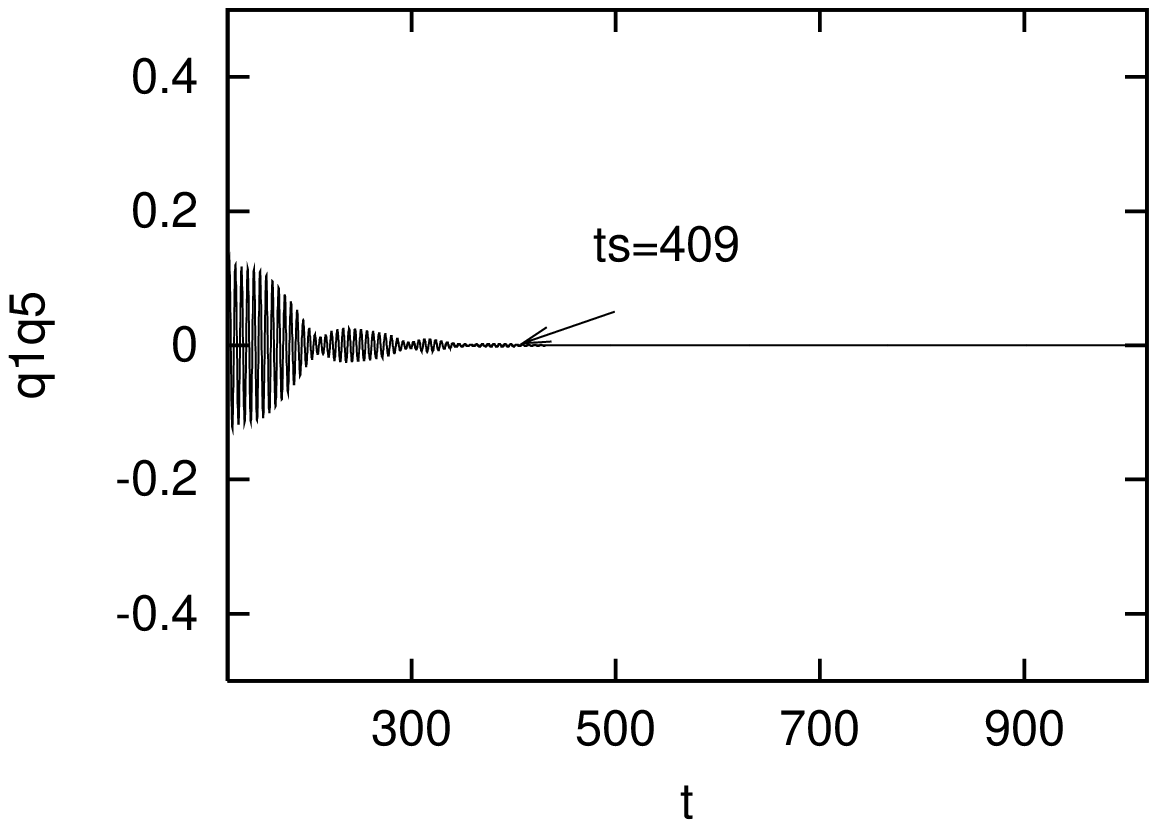}
\includegraphics[width=5cm,height=8cm,angle=-90]{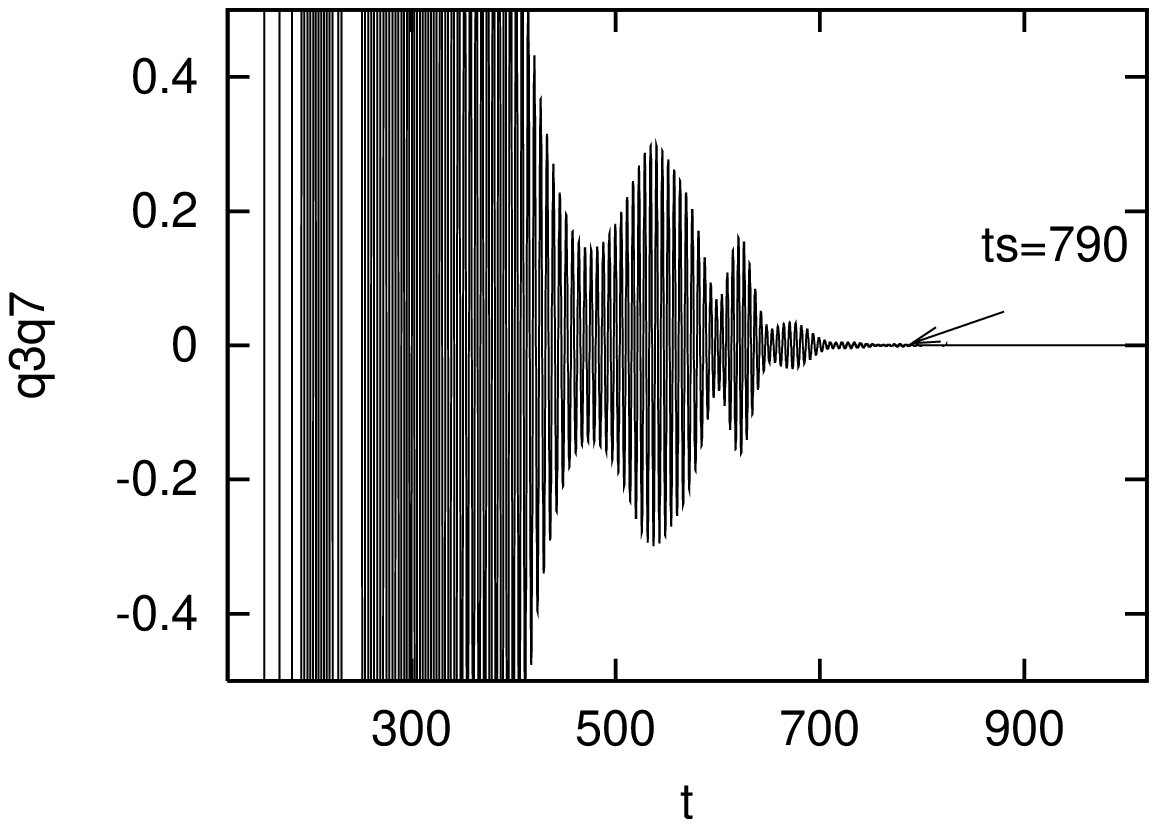}
\end{center}
 \caption{Evolution of $\Delta_{(a,A)}=q_{a}-q_{A}$ and
  $\Delta_{(b,B)}=q_{b}-q_{B}$ {\em vs} $T_{s}$
 for Eqs. (\ref{2aa})--(\ref{2ds}) if $S_{1}=5$ and $S_{2}=0.025$. } 
\label{Fig.3}
\end{figure}
%%%%%%%%%%%%%%%%%%%%%%%%%%%%%%%%%%%%%%%%%%%%%%%%%%%%%%%%%%%%%%%%%%%%%%%%%%%%%%%
 The functions   $\Delta_{(a, A)}= q_{a}-q_{A}$ and $\Delta_{(b,B)}= q_{b}-q_{B}$,  
being a measure of synchronization for the appropriate pairs of  oscillators,
show that the synchronization process for the pair $ ({\bf a},{\bf A})$  is 
faster ($t_{s}^{(a,A)}=409$) than for the pair $({\bf b},{\bf B})$ for which
$T_{s}^{(b,B)}=790$.  Identical results also hold  for
$p_{a}-p_{A}$ and $ p_{b}-p_{B}$.
 A detailed analysis shows that the difference $\Delta (T)=T_{s}^{(b,B)}-T_{s}^{(a,A)}$
  decreases exponentially to zero with increasing  value of
 the interaction parameter
  $\alpha$. Therefore,  the strong linear interaction leads in
  practice to disappearance of the  two-time synchronism. 
 The efficiency of the synchronization process
depends on the values of $\alpha$,
$S_{1}$ and $S_{2}$. By  way of example, this is illustrated in Fig.\ref{Fig.4}, where the
 synchronization times $T_{s}^{(a,A)}$ and $T_{s}^{(b,B)}$  are  presented as
 functions of the control
 parameter $S_{1}$ ( for the fixed values of $\alpha=0.04$ and $S_{2}=5$). 
%%%%%%%%%%%%%%%%%%%%%%%%%%%%%%%%%%% Figure 4 %%%%%%%%%%%%%%%%%%%%%%%%%%%%%%%%%%%%%%%%%%%%%%%%%%%%%%%%%
\begin{figure}
\psfrag{ts2}[B][]{\large $T_{s}^{(a,A)},T_{s}^{(b,B)}$}
\psfrag{s2}[t][]{\large $S_{1}$}
\psfrag{ba}[tc]{$({\bf b},{\bf B})$}
\psfrag{bb}[b]{$({\bf a},{\bf A})$}
\begin{center}
\includegraphics[width=5cm,height=8cm,angle=-90]{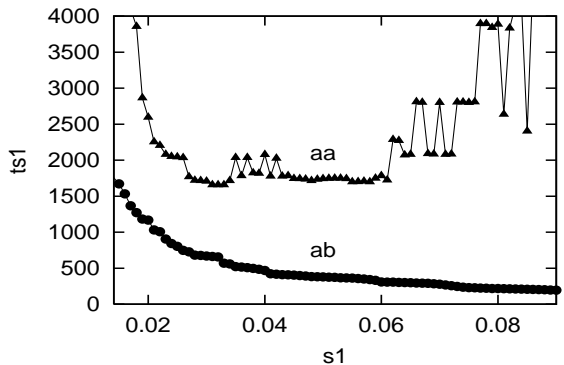} 
\end{center}
\caption{Synchronization times $T_{s}^{(a,A)}$ (triangles) and
  $T_{s}^{(b,B)}$ (bullets)
  {\em vs}  $S_{1}$ if $S_{2}=5$. } 
\label{Fig.4}
\end{figure}
%%%%%%%%%%%%%%%%%%%%%%%%%%%%%%%%%%%%%%%%%%%%%%%%%%%%%%%%%%%
As shown, the pair of oscillators $({\bf a},{\bf A})$  synchronize earlier then the
 pair $({\bf b},{\bf B})$, which is a consequence of  $S_{1}>S_{2}$. 
 If the value $S_{1}$ tends to $S_{2}$, then the difference between 
 $T_{s}^{(A,a)}$ and $T_{s}^{(B,b)}$
vanishes  and finally we observe only one-time synchronization  i.e. $T_{s}^{(a,A)}=T_{s}^{(b,B)}$. 
 The two-time synchronization also occurs for  $S_{1}<0.01$ ( not shown  in
 Fig.\ref{Fig.4} as long synchronization times are  not important in communication). 
 If $S_{1}\rightarrow 0$,  the  times
 $T_{s}^{(a,A)}$ and $T_{s}^{(b,B)}$ tend to infinity. 
 
Let us suppose that in the schematic diagram (Fig.\ref{Fig.1}) only the signal
   from the master subsystem ${\bf A}$ is transmitted to its slave
  counterpart ${\bf a}$. The second signal from ${\bf B}$ to ${\bf b}$
   is turned off (see, Fig.\ref{Fig.1a}). 
%%%%%%%%%%%%%%%%%%  Fig 5 %%%%%%%%%%%%%%%%%%%%%%%%%%%%%%%%%%%%%%%%%%%%%%%%%%%%%%%%%%%%%%%%%%%%%%%
\begin{figure}
\begin{center} 
\includegraphics[width=8cm,height=5cm,angle=0]{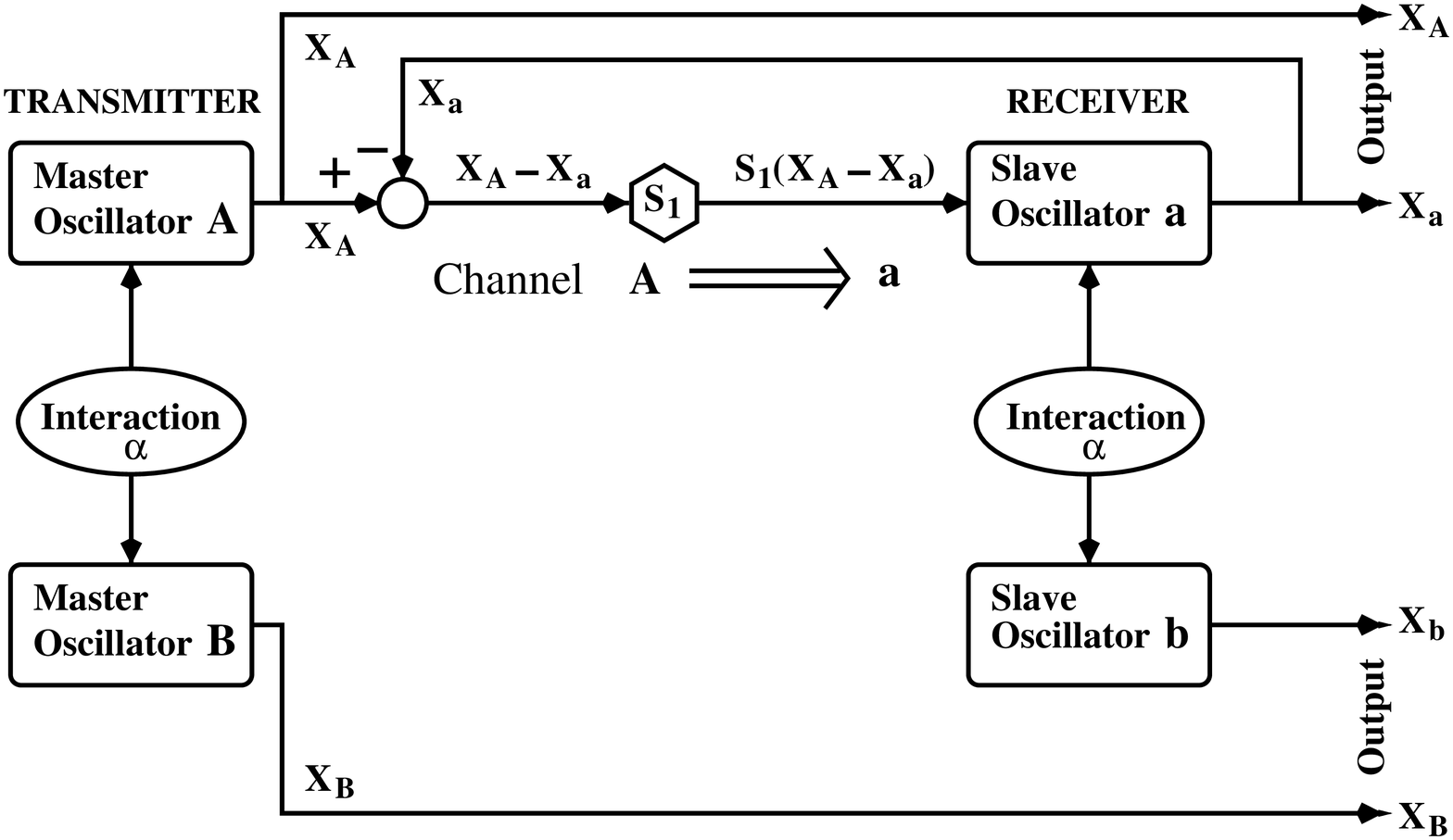} 
\end{center}
\caption{A simplified version of Fig.1. Is it possible to synchronize
  the pair $({\bf b},{\bf B})$ without a signal from the master subsystem  ${\bf
    B}$ to the slave subsystem ${\bf b}$ ?} 
\label{Fig.1a}
\end{figure}
%%%%%%%%%%%%%%%%%%%%%%%%%%%%%%%%%%%%%%%%%%%%%%%%%%%%%%%%%%%%%%%%%%%%%%%%%%%%%%%%%%%%%%%%%%%%%%%%%%%%%
    Numerically, it  means that we solve Eqs.(\ref{2aa})--(\ref{2ds}) for
  $S_{2}=0$  and $S_{1}\neq 0$. The dynamics of synchronization for
   $S_{2}=0$ is studied in the range $0.01<S_{1}<0.1$. The numerical
   analysis shows that  the synchronization times for the pairs
   $({\bf a}, {\bf A})$ and $({\bf b}, {\bf B})$ are approximately equal i.e. 
$T_{s}^{(a,A)}\approx T_{s}^{(b,B)}=T_{s}$. This nearly
   one-time behavior is caused by the small differences
of the values $S_{1}$ and $S_{2}$. 
  Figure \ref{Fig.5} (triangles) presents the synchronization time
 $ T_{s}$ as a function of $S_{1}$.
%%%%%%%%%%%%%%% Figure 5 %%%%%%%%%%%%%%%%%%%%%%%%%%%%%%%%%%%%%%%%%%%%%%%%%%%%%%%%%%%%%%%%%%%%%%%%%%%
\begin{figure}
\psfrag{ts1}[b][]{\large $T_{s}$}
\psfrag{s1}[t][]{\large $S_{1}$}
\psfrag{aa}[bl]{$S_{2}=0$}
\psfrag{ab}[b]{$S_{2}=S_{1}$}
\begin{center}
\includegraphics[width=5cm,height=8cm,angle=-90]{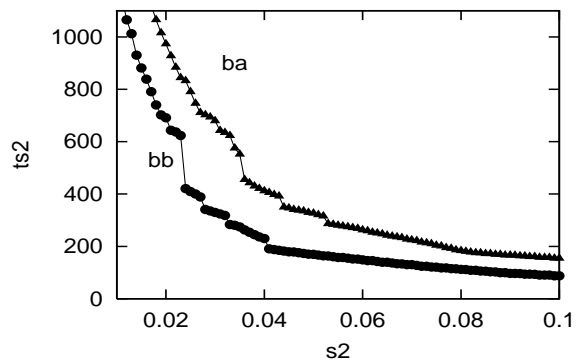} 
\end{center}
\caption{ Synchronization time $T_{s}$
  {\em vs}  $S_{1}$ for   $S_{2}=0$ (triangle) and $S_{2}=S_{1}$ (bullets).  } 
\label{Fig.5}
\end{figure}
%%%%%%%%%%%%%%%%%%%%%%%%%%%%%%%%%%%%%%%%%%%%%%%%%%%%%%%%%%%%%%%%%%%%%%%%%%%%%%%%%%%%%%%%%%%%%%%%%%%%
 The fastest synchronization
  takes place at $S_{1}=0.032$ and the
 synchronization time takes the minimum value $T_{s}=1655$. It is
 interesting to note  that
  in the range $0.041<S_{1}<0.061$
  the synchronization time reaches a nearly constant value. The
   average synchronization time in this region is equal to $<
   T_{s}>=1752$.  As seen from Fig.\ref{Fig.5}, the synchronization process is the most
   effective in the range  $0.025<S_{1}<0.061$ that is
   when $S_{1}\approx \alpha$.
 In the range $0.01<S_{1}$ and $S_{1}>0.1$ the synchronization is not observed. 
 In conclusion, the subsystem ${\bf B}$ does not drive the subsystem ${\bf
  b}$  even though  synchronization of the pair $({\bf b},{\bf B})$ occurs. 
  The synchronization effect is induced
  by the first channel ${\bf A}\Rightarrow {\bf a}$
and  the 
  linear interaction $\alpha$ between the slave oscillators ${\bf a}$
  and ${\bf b}$.
This interaction is a necessary condition for the synchronization process.
 If the channels 
     ${\bf B}\Rightarrow {\bf b}$ and  ${\bf A}\Rightarrow {\bf a}$
   are turned on jointly,  the  dynamics of
   the synchronization presented in Fig.\ref{Fig.5} (bullets) is completely
   modified. For $S_{2}=S_{1}$, the synchronization time $T_{s}$
   decreases exponentially with  increasing values of
   $S_{1}$.  Generally,  the efficiency of the induced synchronization
     ($S_{2}=0,S_{1}\neq 0$)  is always lower than the  synchronization forced by two channels
   ($S_{2}=S_{1}$).
    The occurrence of two-time synchronism and  induced synchronization
  in the dynamical systems presented schematically in
  Figs.\ref{Fig.1}and \ref{Fig.1a} seems  not unique
 and rather common. To observe these effects  we can also  use
  instead of two interacting  Kerr oscillators,  typical 
mechanical systems, for example; the Duffing models considered  in
  \cite{raj2,koz} or other two  high-dimensional systems. However, the effect of induced
  synchronization becomes more and more difficult to achieve 
  if  our  transmitter and receiver  are supplemented by  
 additional oscillators.
 The induced synchronization effect and
  two-time synchronism  are presented for  hyperchaotic beats resembling,
  in a sense, the structure of voice speech.  Induced synchronization seems to 
have  potential application in secure communication  to hide messages \cite{boc}. 
 The appropriate materials useful 
for the generation of beats with chaotic envelopes could be optical
systems consisting of a pair of coupled Kerr fibers
 \cite{kerr}-- \cite{barnet}. 

%%%%%%%%%%%%%%%%%%%%%%%%%%%%%%%%%%%%%%%%%%%%%%%%%%%%%%%%%%%%%%%%%%%%%%%%%%%%%%%%%%%%%%%

\end{document}